\documentclass{article}

\usepackage[eandd, preprint]{neurips_2026}

\usepackage[utf8]{inputenc}   
\usepackage[T1]{fontenc}      
\usepackage{newtxtext,newtxmath} 
\usepackage[hidelinks]{hyperref}  
\usepackage{url}              
\usepackage{booktabs}         
\usepackage{amsfonts}         
\usepackage{amsmath}          
\usepackage{nicefrac}         
\usepackage{microtype}        
\usepackage{xcolor}           
\usepackage{graphicx}         
\usepackage{subcaption}       
\usepackage{multirow}         
\usepackage{enumitem}         

\title{SpikeProphecy: A Large-Scale Benchmark for\\Autoregressive Neural Population Forecasting}

\author{%
  \textbf{John R. Minnick\textsuperscript{1,2,*}, \;
  Jinghui Geng\textsuperscript{2,3}, \;
  Kamran Hussain\textsuperscript{4},} \\[0.35em]
  \textbf{Jesus Gonzalez-Ferrer\textsuperscript{2,5}, \;
  Ash Robbins\textsuperscript{1,2}, \;
  Mohammed A. Mostajo-Radji\textsuperscript{2},} \\[0.35em]
  \textbf{David Haussler\textsuperscript{2,5}, \;
  Jason K. Eshraghian\textsuperscript{1}, \;
  Mircea Teodorescu\textsuperscript{1,2,5}} \\[0.8em]
  \small\textsuperscript{1}Department of Electrical and Computer Engineering, University of California, Santa Cruz, CA, USA \\
  \small\textsuperscript{2}UC Santa Cruz Genomics Institute, University of California, Santa Cruz, CA, USA \\
  \small\textsuperscript{3}Department of Computer Science and Engineering, University of California, Santa Cruz, CA, USA \\
  \small\textsuperscript{4}Department of Applied Mathematics, University of California, Santa Cruz, CA, USA \\
  \small\textsuperscript{5}Department of Biomolecular Engineering, University of California, Santa Cruz, CA, USA
}

\begin{document}
\maketitle
\renewcommand{\thefootnote}{*}%
\footnotetext{Correspondence: \texttt{jrminnic@ucsc.edu}}%
\renewcommand{\thefootnote}{\arabic{footnote}}

\begin{abstract}
Neural population models, which predict the joint firing of many
simultaneously recorded neurons forward in time, are typically
evaluated by a single aggregate Pearson correlation $r$ between
predicted and actual spike counts, a number that masks critical
structure.  We argue that how we evaluate spike forecasting matters
as much as what we build, and introduce SpikeProphecy, the first
large-scale benchmark for causal, autoregressive spike-count
forecasting on real electrophysiology recordings.  Our core contribution is a population metric
decomposition that separates aggregate performance into temporal
fidelity, spatial pattern accuracy, and magnitude-invariant
alignment.  The decomposition surfaces aspects of the underlying
data that an aggregate scalar collapses together.  We
apply the protocol to 105~Neuropixels sessions (Steinmetz~2019 +
IBL Repeated Site; ${\sim}89{,}800$ neurons) with seven architecture
baselines spanning four structural families: four SSMs (three diagonal
and one non-diagonal), a Transformer, an LSTM, and a spiking network.
The decomposition surfaces a brain-region predictability ranking that
reproduces across all seven baselines and survives ANCOVA correction
for firing-statistics constraints (region $\Delta R^{2}{=}0.018$
above the firing-statistics covariates).  It also exposes a
sub-Poisson evaluation floor where rigorous metrics combine with
genuine biophysical constraints on regular spike trains, and yields a
negative result on KL-on-output-rates distillation for ANN$\to$SNN
transfer in this Poisson count domain.
\end{abstract}

\section{Introduction}
\label{sec:intro}

High-density electrode probes such as Neuropixels record the
simultaneous spiking activity of hundreds to thousands of neurons in
behaving animals.  A growing class of sequence models, which we
collectively call \emph{neural population models}, has been trained
on these recordings using modern architectures borrowed from
language and time-series
work~\citep{pandarinath2018inferring, ye2021representation, azabou2023unified}.  These
models are overwhelmingly evaluated on a single downstream task,
\emph{behavioral decoding}: mapping spikes to cursor velocity, stimulus
identity, or movement direction.  The more fundamental computational
challenge is \emph{spike forecasting}: predicting the future firing of
thousands of neurons from their own recent history.  This task matters
for closed-loop brain--computer interfaces (BCIs), where 50--100\,ms
look-ahead predictions compensate for sensing and processing delays, and
for \emph{in silico} neural population simulators (``digital twins'') that
could accelerate BCI algorithm development without animal experimentation.
Despite this importance, \textbf{no established benchmark exists} for
spike-count forecasting at scale on real electrophysiology data.

Standard evaluation practices compound this gap.  The community relies on
aggregate per-neuron Pearson~$r$ as the primary metric, but a single number
hides several axes: brain region differences in predictability, neuron
subpopulation failures, and the distinction between temporal dynamics capture
and spatial pattern fidelity.  An aggregate $r = 0.50$ sounds reasonable,
but decomposing it reveals that temporal population dynamics are
well-captured ($r_\mathrm{pop} = 0.76$) while individual neuron spatial
identity is only moderately captured ($r_\mathrm{spatial} = 0.55$).
We need \emph{population-level metrics} that measure what actually matters
for downstream applications.

We introduce the \textbf{SpikeProphecy benchmark}, designed to address
both gaps simultaneously.  Our contributions are:

\begin{enumerate}[nosep, leftmargin=*]
  \item \textbf{Evaluation protocol (primary contribution).}
    A population-metric decomposition (\texttt{pop\_rate\_r}
    for temporal fidelity, \texttt{spatial\_r} for spatial pattern,
    \texttt{cosine\_sim} for magnitude-invariant alignment) that
    exposes structure invisible to aggregate Pearson~$r$.  The
    decomposition re-orders the brain-region predictability hierarchy
    (\S\ref{sec:regions}), exposes a sub-Poisson evaluation floor
    (\S\ref{sec:subpoisson}), and separates the failure modes of the
    linear-vs.-deep modeling hierarchy
    (\S\ref{sec:arch-results}) that are all missed by single-
    scalar reporting.  We argue the decomposition should become
    standard for neural population forecasting (\S\ref{sec:metrics}).
  \item \textbf{Task \& data.}  The first large-scale autoregressive
    spike-count forecasting benchmark: 105~sessions from two public
    Neuropixels datasets (${\sim}89{,}800$ neurons), with standardized
    preprocessing, temporal splits, and a 14-test data-integrity audit
    suite covering 5~leakage vectors.
  \item \textbf{Seven evaluated architectures.}
    Four SSMs spanning the diagonal/non-diagonal axis (Mamba, HGRN2,
    LRU: diagonal; GatedDeltaNet: non-diagonal), a Transformer,
    an LSTM, and a recurrent spiking network (RSynaptic SNN), all
    trained under identical optimizer, schedule, loss, and data so
    that decomposition findings can be attributed to the task and
    data rather than to any single architecture choice.  Per-architecture
    results are reported in \S\ref{sec:results} and Table~\ref{tab:main-results}.
  \item \textbf{Findings enabled by the benchmark.}  The protocol's
    stratified reporting surfaces structure that standard aggregate
    metrics smooth over: a brain-region predictability ranking that
    survives ANCOVA correction (region $\Delta R^{2}{=}0.018$ above
    firing-statistics covariates), a sub-Poisson evaluation floor
    that combines biophysical hardness with metric harshness, and a
    controlled negative result for KL-on-output-rates distillation
    in this Poisson count setting (we do not claim the result
    generalizes to feature-level or attention-transfer distillation).
  \item \textbf{Public ecosystem.}  Processed tensors and
    preprocessing code (the source recordings remain at their
    public Figshare/IBL repositories), \texttt{pip}-installable
    evaluation toolkit, trained checkpoints, and YAML reproduction
    configs.
\end{enumerate}

\section{Related work}
\label{sec:related}

\paragraph{Neural population models.}
LFADS~\citep{pandarinath2018inferring} uses a sequential VAE to estimate
firing rates but targets latent dynamics inference, not forecasting.
NDT/NDT2~\citep{ye2021representation} applies masked attention to neuronal spiking
for behavioral decoding.  BRAID~\citep{vahidi2025braid} is closest to our
work (input-driven RNNs with a forecasting objective on Steinmetz
recordings) but provides no multi-architecture comparison, no cross-dataset
evaluation, and no population-level metrics.
POCO~\citep{duan2025poco} addresses neural forecasting but on
calcium imaging (continuous fluorescence), not discrete spike counts, making
it a fundamentally different data modality.

\paragraph{Why not foundation models as baselines?}
Recent neural foundation models (POYO/POYO-2, NDT-2, brain2vec) are
not drop-in comparisons: each was pretrained for a different task
(masked-token prediction, latent inference, session-level embeddings)
and would require substantial adaptation to produce causal 50\,ms-bin
rate forecasts (Appendix~\ref{app:foundation-models}).  We acknowledge
this absence as a real gap; our architecture-level findings cannot
claim state-of-the-art.

\paragraph{Neural data benchmarks.}
The Neural Latents Benchmark~(NLB;~\citealt{pei2021neural}) standardized
evaluation of latent variable models using co-smoothing bits per spike
(co-BPS) on 4~motor/cognitive datasets.  NLB targets \emph{latent
inference} (unsupervised rate smoothing), whereas SpikeProphecy targets
\emph{causal forecasting} of future spike counts from history alone.
NeuroBench~\citep{yik2025neurobench} benchmarks neuromorphic algorithms
but on classification tasks (keyword spotting, event vision), not
neural time-series.  SpikeProphecy bridges this gap: the first
benchmark for autoregressive spike prediction at scale, with
evaluation metrics that decompose performance beyond scalar summaries.
Table~\ref{tab:metric-compare} contrasts our evaluation protocol with
existing practice.

\paragraph{Sequence architectures for neural data.}
Structured state-space models (S4, S5) and their successors
(Mamba~\citep{gu2023mamba}, LRU~\citep{orvieto2023resurrecting}) have shown
strong performance on long-range sequence tasks.  We provide the first
systematic comparison of these architectures on neural spike data,
including evaluation beyond Pearson~$r$.

\paragraph{Evaluation methodology.}
Standard neural data metrics (per-neuron $r$, co-BPS, $R^2$) are all
aggregate scalar summaries that collapse temporal and spatial fidelity
into a single number.  HELM~\citep{liang2022holistic} demonstrated in the
LLM domain that multi-axis evaluation (accuracy $\times$ fairness
$\times$ calibration) reveals trade-offs invisible to single-metric
leaderboards.  Our population-level metrics bring this philosophy to
neuroscience, inspired by population vector
analysis~\citep{georgopoulos1986neuronal} but applied to forecast
evaluation.

\paragraph{ANN-to-SNN distillation.}
Knowledge distillation for spiking networks~\citep{hong2025lasnn} has
been studied exclusively on image classification (CIFAR, ImageNet).
No prior work evaluates ANN-to-SNN distillation on spike forecasting;
our negative result is the first empirical demonstration, with a
hypothesized mechanism (the redundancy of soft labels when the target
is already real-valued), for why standard KL-weighted distillation
fails in Poisson regression domains.

\section{Benchmark design}
\label{sec:benchmark}

\subsection{Task formulation}
\label{sec:task}

Given a history window of $T$ spike-count vectors, predict the next
time bin:
\begin{equation}
  X_t = \{\mathbf{x}(t{-}T{+}1), \ldots, \mathbf{x}(t)\}
  \;\longrightarrow\;
  \hat{\mathbf{y}}(t{+}1) \approx \mathbf{x}(t{+}1),
  \quad \mathbf{x}(t) \in \mathbb{Z}_{\geq 0}^{M}
  \label{eq:task}
\end{equation}
where $M$ is the number of simultaneously recorded neurons in a
session (varies between sessions, up to $M_\mathrm{max}{=}1{,}998$
in the largest IBL session), $\Delta t = 50$\,ms, and $T = 10$
bins (500\,ms history).  Sessions are zero-padded to $M_\mathrm{max}$
with a per-sample binary channel mask; the loss operates only on
real channels.  The model outputs softplus rates $\hat{\lambda}_i > 0$
trained with Poisson NLL.  Two constraints define the task:
\emph{strictly autoregressive} (intrinsic covariates only: past
spikes, no stimulus/behavioral features) and \emph{causal} (no
future context at any layer).  This rules out stimulus information
unavailable at BCI inference time and the bidirectional attention
used in masked-token foundation models like POYO/NDT-2.

\subsection{Datasets}
\label{sec:datasets}

\paragraph{Steinmetz 2019 (39~sessions).}
Ten mice, 39~recording sessions using Neuropixels probes spanning
visual cortex, motor cortex, hippocampus, thalamus, and
midbrain~\citep{steinmetz2019distributed}.  Up to 1{,}240 simultaneously
recorded neurons per session.  Visual discrimination task
(${\sim}$2-hour recordings).  Single lab, single rig, controlled
conditions.  Source data are publicly available on Figshare under
CC-BY-4.0 (\url{https://doi.org/10.6084/m9.figshare.9598406.v2});
our processed 50\,ms-binned tensors plus per-session metadata
(split boundaries, brain-region labels) are released as a
HuggingFace dataset
(\url{https://huggingface.co/datasets/mysteriousauthor/spikeprophecy-steinmetz}).

\paragraph{IBL Repeated Site (66~sessions).}
Multi-lab international consortium, standardized probe trajectory
across 9 labs~\citep{international2025reproducibility}.  Up to 1{,}998
simultaneously recorded neurons.  Same task paradigm across different
labs, mice, and rigs; tests cross-lab generalization.  Source data
are publicly available via the IBL Open Neurophysiology Environment
ONE API (\url{https://www.internationalbrainlab.com/data}); as
above, we redistribute only the processed tensors and the
preprocessing pipeline.

\paragraph{Processing and splits.}
Raw spike times are binned at 50\,ms into integer count vectors.
Each session is split temporally into three contiguous blocks
(70/15/15 train/val/test, ordered first/middle/last in time) to
prevent information leakage.  Interleaved or random splits would
introduce trivial leakage on autoregressive forecasting because
adjacent 50\,ms bins are temporally correlated.

\paragraph{Auditable leakage suite.}
We ship a 14-test audit suite that runs on every commit and
verifies five leakage vectors (Appendix~\ref{app:leakage-suite}).
The Population-GLM result in Table~\ref{tab:main-results}
($r{=}1.000$ on train, $r{=}{-}0.015$ on val) is the canonical
catch.

\subsection{Evaluation protocol: population metric decomposition}
\label{sec:metrics}

Each of the three metrics below is individually well-precedented
(population vector analysis~\citep{georgopoulos1986neuronal},
per-bin cross-neuron correlation, magnitude-invariant cosine
alignment); our contribution is standardizing the triplet as the
reporting protocol for spike-count forecasting and showing
empirically (\S\ref{sec:findings}) that it surfaces structure
aggregate~$r$ hides.  The three describe distinct axes (one
marginalizes neurons, one marginalizes time, one removes
magnitude); they are not orthogonal components of an algebraic
identity summing to Wt-$r$.

\paragraph{Standard metrics (baseline protocol).}
Existing benchmarks rely on aggregate Pearson~$r$
(\emph{weighted Pearson~$r$}, Wt-$r$), per-neuron median~$r$,
co-smoothing bits per spike (co-BPS), Poisson NLL, and mean
absolute error (MAE) -- all scalar summaries that conflate
temporal and spatial fidelity (Appendix~\ref{app:metric-compare}
contrasts the protocols).  We define Wt-$r$ as the
activity-weighted mean of per-neuron Pearson correlations:
\begin{equation}
  \mathrm{Wt}\text{-}r = \frac{\sum_{i=1}^{M} w_i \cdot
      \mathrm{Pearson}\bigl(y_i,\hat y_i\bigr)}
      {\sum_{i=1}^{M} w_i},
  \quad w_i = \mathrm{Var}\bigl(y_i\bigr),
  \label{eq:wt-r}
\end{equation}
where weighting downweights near-silent neurons.  Wt-$r$ is the
headline metric in Table~\ref{tab:main-results}, reported for
backwards comparability.

\paragraph{Population metric decomposition (our contribution).}
Let $\mathbf{y}(t), \hat{\mathbf{y}}(t) \in \mathbb{R}^M$ denote
ground-truth and predicted activity vectors at time bin~$t$
across $M$ neurons and $T_\mathrm{eval}$ evaluation bins.

\textbf{(i) Population Rate $r$ (\texttt{pop\_rate\_r}).}
Temporal fidelity: \emph{when is the population active?}
\begin{equation}
  r_\mathrm{pop} = \mathrm{Pearson}\!\left(
    \Big[{\textstyle\sum_{i=1}^{M} y_i(t)}\Big]_{t=1}^{T},\;
    \Big[{\textstyle\sum_{i=1}^{M} \hat{y}_i(t)}\Big]_{t=1}^{T}
  \right)
  \label{eq:pop-rate}
\end{equation}
Marginalizes over neuron identity to measure ensemble rate
envelope tracking.

\textbf{(ii) Spatial Pattern $r$ (\texttt{spatial\_r}).}
Spatial fidelity: \emph{which neurons fire?}
\begin{equation}
  r_\mathrm{spatial} = \frac{1}{T_\mathrm{eval}}
  \sum_{t=1}^{T_\mathrm{eval}}
  \mathrm{Pearson}\!\big(\mathbf{y}(t),\;\hat{\mathbf{y}}(t)\big)
  \label{eq:spatial}
\end{equation}
Per-timebin cross-neuron correlation, capturing identification
of the active subset at each moment.

\textbf{(iii) Population Cosine Similarity (\texttt{cosine\_sim}).}
Magnitude-invariant pattern alignment: \emph{relative activation
regardless of overall rate?}
\begin{equation}
  \mathrm{cos\_sim} = \frac{1}{T_\mathrm{eval}}
  \sum_{t=1}^{T_\mathrm{eval}}
  \frac{\mathbf{y}(t) \cdot \hat{\mathbf{y}}(t)}
       {\|\mathbf{y}(t)\| \;\|\hat{\mathbf{y}}(t)\|}
  \label{eq:cosine}
\end{equation}
Normalizing magnitudes isolates pattern fidelity from rate
calibration.  Because spike counts and softplus predictions are
non-negative, cosine here lives in the positive orthant and has
a non-zero floor: train-set-mean reaches $\cos{=}0.31$
(empirical floor), the autoregressive GLM $0.24$, the
modern-recurrence cluster $0.61\text{--}0.63$, the trailing
baselines $\cos{\approx}0.58$ (Table~\ref{tab:main-results}).
The dynamic range $0.31\text{--}0.63$ remains discriminative.

We additionally advocate reporting results \emph{stratified} by
brain region (\S\ref{sec:regions}) and Fano factor class
(\S\ref{sec:subpoisson}).

\subsection{Architecture baselines}
\label{sec:architectures}

All architectures share input pipeline, loss, optimizer, and
training schedule (50~epochs, AdamW, cosine LR with warmup,
seed=42; full per-arch hyperparameters in
Appendix~\ref{app:arch-hparams}).

\begin{table}[!ht]
  \caption{Architecture baselines in the SpikeProphecy benchmark.
    All models use shared output heads and identical training
    hyperparameters.}
  \label{tab:architectures}
  \centering
  \small
  \begin{tabular}{@{} l l c l @{}}
    \toprule
    \textbf{Architecture} & \textbf{Type} & \textbf{Params}
      & \textbf{Key property} \\
    \midrule
    Mamba     & Diagonal selective SSM & 1.95M & Input-dependent gating, $O(T)$ \\
    HGRN2     & Diagonal gated linear RNN & 1.82M & State expansion, $O(T)$ \\
    LRU    & Diagonal linear recurrence & 1.23M & Ring eigenvalue init \\
    GatedDeltaNet & Non-diagonal delta-rule SSM & 1.43M & Matrix state per head, $O(T)$ \\
    Transformer & Causal attention     & 2.22M & Global context, $O(T^2)$ \\
    LSTM      & Gated recurrence       & 2.22M & Classical baseline \\
    SNN       & Spiking (RSynaptic)    & 965K  & Event-driven, neuromorphic (3L) \\
    \midrule
    Autoreg GLM    & Poisson, own $T$-step hist. & ${\sim}10/N$  & No cross-neuron info \\
    Population GLM & Poisson, full $(T, M)$ hist. & ${\sim}7\text{K}/N$ & Linear pop baseline \\
    \bottomrule
  \end{tabular}
\end{table}

\begin{figure*}[!t]
  \centering
  \includegraphics[width=\textwidth]{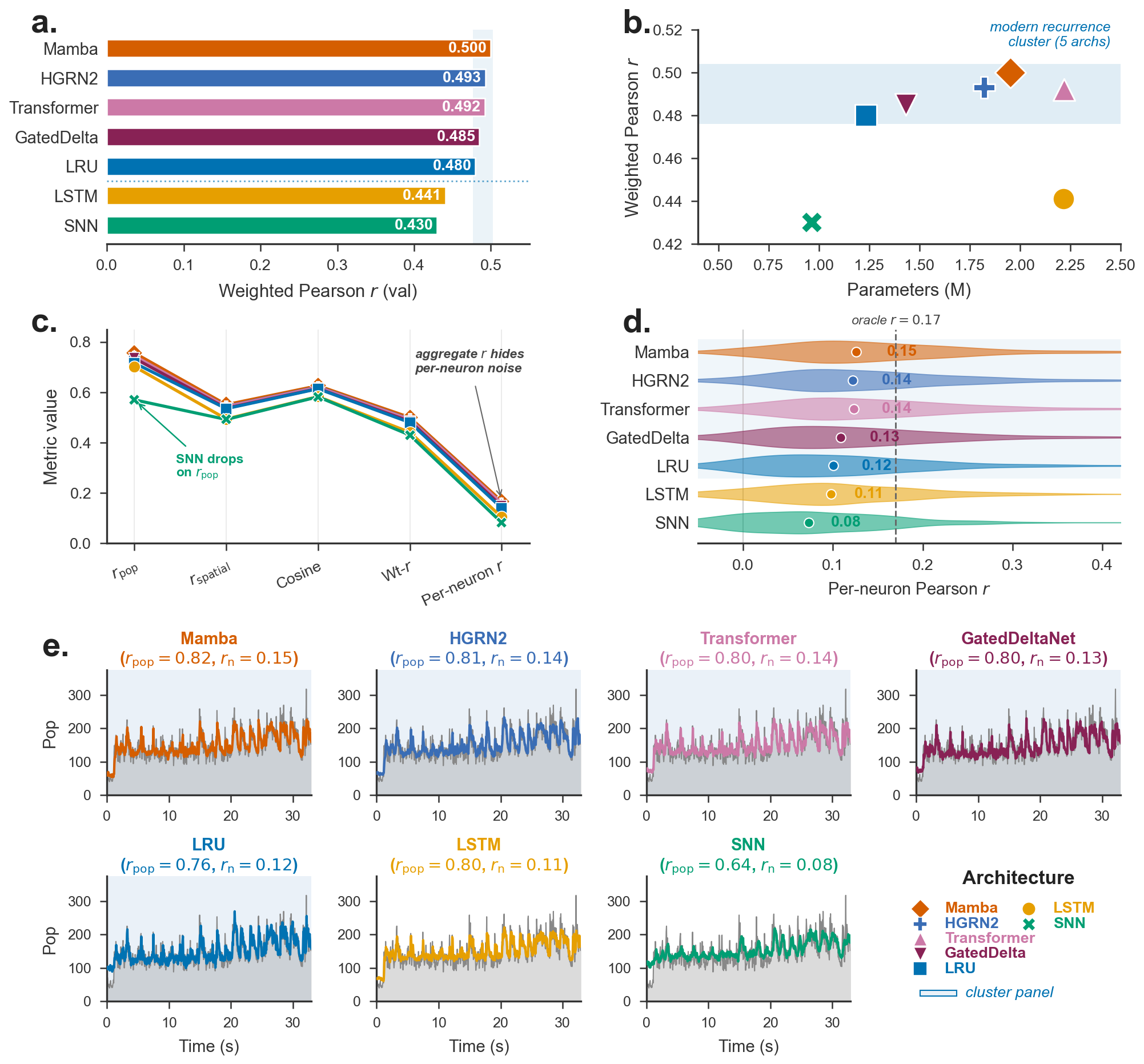}
  \caption{\textbf{SpikeProphecy benchmark overview.}  Per-arch
    color/marker is consistent across (a)--(e).  (a)--(d) summarize
    all 39 Steinmetz sessions ($27{,}212$ neurons); (e) shows
    predictions on near-median session~4.
    \textbf{(a)}~Absolute Wt-$r$, sorted: five cluster baselines
    fall in $0.480\text{--}0.500$ (blue band), LSTM $0.441$, SNN
    $0.430$; dotted divider marks the Wilcoxon-significant
    cluster-vs.-LSTM/SNN gap ($p{<}10^{-7}$).
    \textbf{(b)}~Pareto: Wt-$r$ vs.\ parameter count.
    \textbf{(c)}~Decomposition trace across five normalized metrics:
    cluster archs overlap; SNN drops on $r_\mathrm{pop}$;
    per-neuron~$r$ collapses an order of magnitude below the
    aggregate metrics for every architecture.
    \textbf{(d)}~Per-neuron $r$ violins (${\sim}700$ neurons of
    session~4).  Empirical oracle ceiling ($r{=}0.17$, dashed).
    \textbf{(e)}~Population-rate small multiples on a near-median
    Steinmetz session ($N{=}703$, ${\sim}33$\,s); cluster panels
    carry a faint blue background; titles report session-4
    $r_\mathrm{pop}$ and per-neuron $r_\mathrm{n}$.}
  \label{fig:hero}
  \label{fig:pareto}
  \label{fig:radar}
\end{figure*}


\section{Results}
\label{sec:results}

\subsection{Decomposition reveals structure invisible to aggregate $r$}
\label{sec:findings}

The population metric decomposition and stratified evaluation reveal
two findings that are \textbf{invisible to standard aggregate
metrics}.  Each demonstrates structure that would be missed by
reporting a single Pearson~$r$.

\begin{figure*}[!ht]
  \centering
  \includegraphics[width=\textwidth]{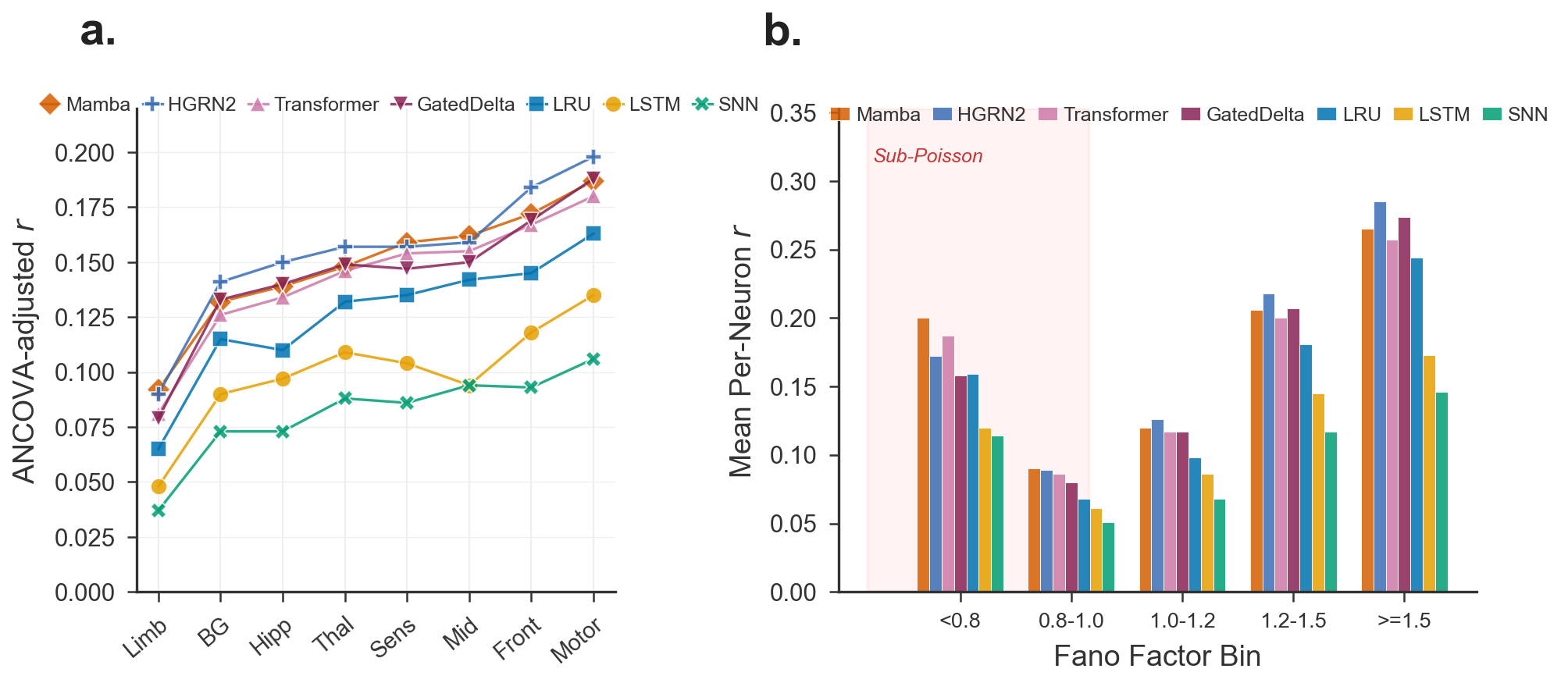}
  \caption{\textbf{Findings enabled by the population metric decomposition.}
    (a)~ANCOVA-adjusted per-neuron $r$ across 8 functional brain
    regions (39-session aggregate, 27{,}212 neurons), one line per
    architecture.  Covariates: log mean firing rate + Fano factor.
    All seven baselines trace the same monotonic hierarchy across
    regions despite their absolute-level differences, indicating the
    predictability ranking is a property of the data, not of any
    single architecture.  Motor cortex and midbrain are easiest;
    Limbic/Other and Hippocampal are hardest.  Cluster archs occupy
    the upper band (Mamba, HGRN2, Transformer, GatedDeltaNet, LRU);
    LSTM and SNN sit visibly below at every region.
    (b)~Per-neuron $r$ by Fano factor bin across all seven baselines
    (39-session aggregate, val split, 26{,}631 neurons binned).  The
    0.8\,$\leq$\,FF\,$<$\,1.0 sub-Poisson bin is the lowest for
    \emph{every} architecture, a fundamental evaluation ceiling that
    standard aggregate $r$ conceals.}
  \label{fig:findings}
\end{figure*}

\paragraph{Finding 1: A functional brain-region predictability hierarchy.}
\label{sec:regions}
Parsing Allen CCF brain region labels for 27{,}212 neurons across
39~sessions reveals a predictability hierarchy that remains
significant after ANCOVA correction for log firing rate and Fano
factor (Figure~\ref{fig:findings}a).  The full ANCOVA model
(\texttt{model\_r} $\sim$ \texttt{log\_rate} $+$ \texttt{fano} $+$
\texttt{region}) reaches $R^{2}{=}0.275$ at the 8-system grouping; the
covariates alone reach $R^{2}{=}0.257$, so the \emph{region-incremental}
contribution beyond firing-statistics is $\Delta R^{2}{=}0.018$ at this
granularity.  The increment is modest but reproducible: it survives
ANCOVA correction (partial $F{>}10^{2}$, $p < 10^{-77}$) and is
robust to grouping choice.  At a coarser 4-class grouping
(Cortex/Subcortex/Hippocampal/Other) the increment drops to
$\Delta R^{2}{=}0.011$; at the fine 54-region Allen acronym level it
rises to $\Delta R^{2}{=}0.053$ (Appendix~\ref{app:region-sensitivity}).
As Figure~\ref{fig:findings}a shows, the ranking itself is
reproducible across all seven baselines: every architecture traces
the same monotonic hierarchy across the eight functional regions
despite their absolute-level differences, which is the more
substantive finding than the absolute $R^{2}$.  The
uncorrected Kruskal--Wallis statistic ($H{=}1{,}056$, $p < 10^{-200}$)
is inflated by the large neuron count and should not be interpreted
as an effect size.

\begin{table}[!ht]
  \caption{\textbf{Finding 1: Brain-region predictability hierarchy}
    (Mamba per-neuron Pearson~$r$, val split; 21{,}689 of 27{,}212
    Steinmetz neurons assigned to one of 8 Allen CCF functional
    systems).  This ranking is hidden by aggregate $r$ and
    persists after ANCOVA correction for firing rate and Fano factor
    (Figure~\ref{fig:findings}a).
    $\bar{r}$: mean per-neuron Pearson~$r$;
    \%sub-P: fraction with Fano factor~$< 1$.}
  \label{tab:regions}
  \centering
  \small
  \begin{tabular}{@{} l r c c @{}}
    \toprule
    \textbf{Region group} & \textbf{$N$}
      & \textbf{Mean $r$}
      & \textbf{\%sub-P} \\
    \midrule
    Motor cortex          & 1{,}780 & \textbf{0.175} & 25\% \\
    Thalamus              & 5{,}240 & 0.170 & 11\% \\
    Midbrain/Brainstem    & 3{,}440 & 0.168 & 42\% \\
    Sensory cortex        & 2{,}838 & 0.156 & 28\% \\
    Frontal/Association   & 2{,}507 & 0.153 & 34\% \\
    Hippocampal           & 3{,}736 & 0.130 & 23\% \\
    Basal ganglia         & 1{,}689 & 0.121 & 49\% \\
    Limbic/Other          &    459 & 0.080 & 22\% \\
    \bottomrule
  \end{tabular}
\end{table}

This hierarchy \textbf{survives} ANCOVA controlling for log firing rate
and Fano factor ($R^{2}{=}0.275$, $\Delta R^{2}{=}0.018$ above
the firing-statistics covariates, $p < 10^{-77}$), supporting its
interpretation as a property of the regions themselves rather than
of firing statistics alone.  The ranking is also stable across the
seven baselines we tested (Mamba, HGRN2, Transformer, GatedDeltaNet,
LRU, LSTM, SNN); we frame this as a within-recipe consistency
check rather than as evidence of architecture-independence in the
strong sense, since these baselines share an optimizer, schedule,
loss, and training data.  Stimulus-driven regions (motor, midbrain)
exhibit highly predictable 50\,ms dynamics, while memory/association
regions (hippocampus) have internally-generated dynamics operating on
longer timescales than our 500\,ms context, identifying a concrete
target for future model development.

\paragraph{Finding 2: A sub-Poisson evaluation floor with two
compounding causes.}
\label{sec:subpoisson}
28\% of neurons have Fano factor $< 1$ (more regular than Poisson).
These regular-firing neurons are the \textbf{hardest} to predict
across \emph{all} architectures tested (mean $r = 0.073$ vs.\ $0.151$
for super-Poisson neurons; Figure~\ref{fig:findings}b).  Part of this
gap is biophysical (the model lacks the relevant within-neuron
oscillator covariate) and part is a metric artifact (Pearson punishes
timing jitter quadratically in low-variance regimes); the two
contributions are entangled in this measurement.  Why are
clock-like neurons so hard to forecast?  Two compounding reasons.
First, sub-Poisson regularity reflects intrinsic oscillatory or
pacemaker dynamics driven by single-cell biophysics rather than by
the population history available to the model on a 50\,ms timescale,
so the model lacks the relevant covariate.  Second, when a neuron's
true count distribution has very low variance, even small phase
mispredictions translate into large per-bin Pearson penalties: the
metric punishes timing jitter quadratically in a low-variance regime.  These neurons therefore
impose a \textbf{hard noise floor} on aggregate metrics:
${\sim}0.07$ regardless of model quality.  This has a concrete
reporting consequence: without Fano-stratified evaluation, a model
that improves only on super-Poisson neurons (where the signal is)
will look marginal on aggregate~$r$ even if its real-signal
improvement is substantial.


\subsection{Baselines and the linear bracket}
\label{sec:arch-results}

The findings above were generated by running the seven architecture
baselines and two GLM controls under identical training conditions.
We report their relative performance here for completeness; the
ranking is a consequence of the protocol, not a contribution of the
paper.  Table~\ref{tab:main-results} gives full per-architecture
results on Steinmetz~2019 (39~sessions, 27{,}212~neurons).

\begin{table}[!ht]
  \caption{Per-architecture results on Steinmetz~2019
    (39~sessions, 27{,}212 neurons, val split).  Five
    modern-recurrence baselines cluster at
    $r{=}0.480\text{--}0.500$; LSTM and SNN fall below.  Four
    non-deep baselines anchor the floor: \emph{persistence} and
    \emph{train-mean} calibrate, the autoregressive GLM and
    CV-ridge population GLM bracket the linear regime, and the
    fixed-$\alpha$ population GLM is included as a leakage-audit
    catch ($r{=}{-}0.015$ on val).  Cluster-vs.-trailing
    significance (Wilcoxon, $p{<}10^{-7}$), the
    no-within-cluster-ordering caveat, and per-session SEs are
    detailed in \S\ref{sec:arch-results} and
    Appendix~\ref{app:per-session-se}.  CV-ridge MAE is inflated
    by the softplus link
    (Appendix~\ref{app:cv-ridge-glm}).}
  \label{tab:main-results}
  \centering
  \small
  \setlength{\tabcolsep}{4pt}
  \begin{tabular}{@{} l r c c c c c @{}}
    \toprule
    \textbf{Model} & \textbf{Params}
      & \textbf{Wt-$r$$\uparrow$}
      & $\boldsymbol{r_\mathrm{pop}}$$\uparrow$
      & $\boldsymbol{r_\mathrm{spat}}$$\uparrow$
      & \textbf{cos}$\uparrow$
      & \textbf{MAE}$\downarrow$ \\
    \midrule
    \textbf{Mamba}
      & 1.95M & \textbf{0.500} & \textbf{0.756} & \textbf{0.551}
      & \textbf{0.626} & \textbf{0.283} \\
    HGRN2
      & 1.82M & 0.493 & 0.740 & 0.544 & 0.621 & 0.286 \\
    Transformer
      & 2.22M & 0.492 & 0.744 & 0.543 & 0.620 & 0.286 \\
    GatedDeltaNet
      & 1.43M & 0.485 & 0.735 & 0.537 & 0.615 & 0.288 \\
    LRU
      & 1.23M & 0.480 & 0.716 & 0.535 & 0.614 & 0.290 \\
    LSTM
      & 2.22M & 0.441 & 0.702 & 0.494 & 0.583 & 0.298 \\
    SNN (3L)
      & 965K  & 0.430 & 0.570 & 0.492 & 0.582 & 0.301 \\
    \midrule
    Persistence
      & 0     & $-$0.003 & 0.030 & 0.024 & 0.114 & 0.260 \\
    Train-mean
      & ${\sim}1/N$ & 0.000 & 0.000 & 0.069 & 0.310 & 0.350 \\
    Autoreg GLM
      & ${\sim}10/N$   & $+$0.001 & 0.150 & 0.091 & 0.237 & 0.253 \\
    Pop GLM (fixed $\alpha$, leakage catch)
      & ${\sim}7\mathrm{K}/N$ & $-$0.015 & -- & -- & -- & -- \\
    Pop GLM (CV-ridge $\alpha$)
      & ${\sim}7\mathrm{K}/N$ & $+$0.023 & 0.143 & 0.076 & 0.224 & 0.813 \\
    \bottomrule
  \end{tabular}
\end{table}


All seven baselines surface the same structural findings under our
shared training recipe (region hierarchy, Fano stratification);
this is a within-recipe consistency check, not a strong
architecture-independence claim (\S\ref{sec:related}).

The five modern-recurrence architectures (Mamba, HGRN2, Transformer,
GatedDeltaNet, LRU) cluster within 2~percentage points
($r{=}0.480\text{--}0.500$).  The within-cluster spread is on the
order of per-session noise on this 39-session draw
(Appendix~\ref{app:per-session-se}), so we report the cluster as a
band rather than an ordering.  The classical LSTM ($r{=}0.441$) and
depth-matched SNN ($r{=}0.430$) fall below the cluster; the SNN
ablation (Appendix~\ref{app:snn-distill}) recovers to $r{=}0.481$ at
1~layer with $36\%$ of the parameters.

\textbf{Trivial baselines calibrate the metric scale.}  Two
parameter-free baselines anchor the leaderboard at the bottom.
\emph{Persistence} (predict the previous bin) reaches Wt-$r{=}{-}0.003$
and $r_\mathrm{pop}{=}0.03$, confirming that 50\,ms own-bin history
carries essentially no information about the next bin.
\emph{Train-mean} (predict the per-neuron training-set mean, a constant
in time) achieves Wt-$r{=}0$ and $r_\mathrm{pop}{=}0$ by construction
but reaches $\cos{=}0.31$, which is the true positive-orthant floor for
cosine on non-negative spike-count vectors; deep models reaching
$\cos{=}0.62\text{--}0.63$ doubles this floor.  Persistence achieves
the lowest MAE in the table (0.260, below Mamba's 0.283), because most
50\,ms bins contain zero spikes and persistence predicts zero on those
bins; this confirms that MAE on this task is dominated by the rate
floor and should be read alongside Pearson-based metrics rather than
as a primary objective.

\textbf{Linear baselines fail for distinct reasons; deep models
succeed.}  Two controlled GLM baselines bracket what a linear model
can do on this task.  The per-neuron autoregressive GLM (each neuron's
own $T{=}10$-bin history, 10 features per model) reaches
$r \approx 0.001$ on held-out time despite fitting training data at a
modest level ($r \approx 0.17$ on a representative session), indicating
that 50\,ms own-history dynamics do not generalize across the temporal
split; firing statistics are non-stationary within a 2-hour recording.
The population GLM (full flattened $(T, M)$ history, ${\sim}7\text{K}$
features per neuron, $\alpha{=}10^{-4}$) achieves $r = 1.000$ on train
but $r = -0.015$ on val.  We include this configuration deliberately as
the canonical leakage-suite catch (\S\ref{sec:datasets}): a feature-rich
linear model on a temporally-split benchmark is guaranteed to overfit
under fixed regularization, and the audit suite is what flags it.  A
fairer linear comparison is the same architecture with $\alpha$ tuned
per session on val (grid: $\{10, 10^{2}, 10^{3}, 10^{4}\}$, picked by
val mean per-neuron $r$).  This CV-ridge variant escapes the
catastrophic overfit (Wt-$r{=}{+}0.023$ on val, vs.\ $-0.015$ at fixed
$\alpha$) but still falls far short of deep models: per-session val
mean $r{=}0.027 \pm 0.005$, comparable to the autoregressive GLM and
${\sim}20\times$ below the deep architectures.  The CV-ridge population
metrics ($r_\mathrm{pop}{=}0.14$, $r_\mathrm{spatial}{=}0.08$,
$\cos{=}0.22$) sit in the same regime as the autoregressive GLM, not
the deep cluster.  Full per-session results in
Appendix~\ref{app:cv-ridge-glm}.  The finding here is a methodological
one: all seven deep architectures remain stable under the same
temporal split where the linear population model collapses or only
marginally clears zero, achieving $3\text{--}5\times$ gain over the
autoregressive baseline.



\subsection{Cross-dataset scaling on the Mamba teacher (summary)}
\label{sec:scaling}

The Mamba teacher trained on the 105-session combined Steinmetz+IBL
corpus reaches Wt-$r{=}0.556$ vs.\ $0.500$ on Steinmetz-only
($+11\%$), with $r_\mathrm{pop}$ and cosine gaining similarly and
spatial~$r$ slightly degrading; cross-lab transfer required no
architectural modification.  Full table and caveats (confound
between session count and channel-space size, IBL-only vs.\ combined
qualitative comparison) are in Appendix~\ref{app:scaling}.

\subsection{SNN baselines and teacher distillation (summary)}
\label{sec:snn-summary}

The best standalone SNN (1~layer, 702K~parameters; $36\%$ of
Mamba's count) matches $96\%$ of Mamba's Wt-$r$ and $91\%$ of its
cosine; adding depth \emph{hurts} and no KL-on-output-rates
distillation variant ($\beta \in \{0, 0.5, 1.0\}$) beats standalone
training.  We attribute this to teacher rates carrying no ``dark
knowledge'' advantage when targets are already real-valued.  Full
depth and $\beta$ ablations + mechanism discussion in
Appendix~\ref{app:snn-distill}; the scope-claim is
\emph{output-distillation only} (feature-level / attention transfer
not tested).

\section{Discussion}
\label{sec:discussion}

\paragraph{Summary of findings.}
The protocol's three metrics plus stratified reporting (by region
and Fano factor), applied to the 39-session Steinmetz benchmark with
seven architecture baselines, surfaces four findings that
traditional single-scalar correlation reporting would obscure:
\begin{enumerate}[nosep, leftmargin=*, topsep=0pt]
  \item \textbf{Brain-region predictability ranking, stable across
    the seven baselines we tested} across 8~Allen functional systems
    (region $\Delta R^{2}{=}0.018$ above firing-statistics covariates,
    full-model $R^{2}{=}0.275$, $p{<}10^{-77}$); ANCOVA \emph{re-orders}
    the ranking relative to raw means.
  \item \textbf{Sub-Poisson evaluation floor}: $28\%$ of neurons are
    more regular than Poisson and contribute mean $r{\approx}0.07$
    regardless of model.  The gap is partly biophysical (missing
    oscillator covariate) and partly metric-artifactual (Pearson
    punishes timing jitter in low-variance regimes).
  \item \textbf{Linear-vs.-deep modeling hierarchy} with distinct
    failure modes: per-neuron autoregressive GLM
    ($r{\approx}0$, non-stationary), population GLM
    ($r{=}{-}0.015$, capacity-overfit), deep models ($r{\sim}0.5$,
    succeed nonlinearly).
  \item \textbf{KL-on-output-rates distillation does not help in
    this Poisson count domain} (negative result, \S\ref{sec:snn-summary},
    Appendix~\ref{app:snn-distill}).  The standalone 1L SNN matches
    96\% of the cluster-leader Wt-$r$ at $36\%$ of the parameters; KL
    on softplus rates degrades performance.
\end{enumerate}

\paragraph{Use cases and implications.}
Decomposed metrics expose deployment-relevant trade-offs aggregate~$r$
collapses: high cosine + mediocre aggregate~$r$ suits brain-state
classification or coarse population-vector decoding, while strong
$r_\mathrm{pop}$ + weak $r_\mathrm{spatial}$ suits population-rate
tracking (seizure detection, BCI gating) but not single-neuron
stimulation.  The hippocampal gap ($r = 0.10$) identifies a concrete
target for extended-context or latent-state models, and the
sub-Poisson floor motivates dispersion-aware losses (e.g.,
Conway--Maxwell--Poisson).  A pre-registered diagonal-vs.\
non-diagonal SSM test was inconclusive
(Appendix~\ref{app:diagonal-ssm}).  We propose SpikeProphecy as a
reporting standard, analogous to HELM for LLMs or NLB for latent
dynamics.

\paragraph{Limitations.}
(1)~One-step ($H{=}1$), 50\,ms bins.  (2)~No hardware-deployed
neuromorphic evaluation, no behavioral-decoding downstream task.
(3)~Steinmetz + IBL only (mouse Neuropixels, overlapping
visual/decision tasks), so cross-dataset claims are
\emph{cross-lab within paradigm}.  (4)~No foundation-model
baselines (POYO, NDT-2, brain2vec; see \S\ref{sec:related}).
(5)~Three architectures are 3-seed, four are single-seed; seed
variance is two orders of magnitude below per-session SE so the
39-session sample is the within-cluster bottleneck, but the
cluster-vs.-LSTM/SNN gap survives both
(Appendix~\ref{app:per-session-se}).

\section{Release}
\label{sec:release}

We release processed 50\,ms-binned tensors
(\url{https://huggingface.co/datasets/mysteriousauthor/spikeprophecy-steinmetz},
CC-BY-4.0; IBL at acceptance), a \texttt{pip install spikeprophecy}
evaluation toolkit (MIT), trained checkpoints for all seven
architectures, and YAML reproduction configs.  Maintenance plan and
source-data URLs in Appendix~\ref{sec:datasheet}.

\paragraph{Broader impacts.}
Standardized evaluation infrastructure lowers entry cost for ML
researchers entering computational neuroscience and the
leakage-audit suite establishes a reproducibility floor.
Potential negative use: spike-forecasting models could be
deployed in invasive neural interfaces without sufficient
additional validation; we explicitly disclaim clinical
suitability.  See the Croissant \texttt{rai:dataSocialImpact}
field for the machine-readable version.


{\small
\bibliographystyle{plainnat}
\bibliography{spikeProphecy_reference}
}

\appendix

\section{Additional results}
\label{sec:appendix}

\subsection{Why foundation-model baselines are not included}
\label{app:foundation-models}

POYO/POYO-2~\citep{azabou2023unified} tokenize individual spike events
and pretrain for \emph{masked-token prediction} across heterogeneous
sessions; adapting POYO to produce a causal, 50\,ms-bin rate
forecast requires swapping its output head and reintroducing a
temporal causality constraint that its bidirectional attention
violates.  NDT-2~\citep{ye2021representation} targets \emph{latent inference}
(unsupervised smoothing of current-bin rates under co-smoothing) and
is evaluated primarily on behavioral decoding; forecasting the next
bin from history alone is not a supported task in the public
release.  brain2vec and similar representation-learning approaches
produce session-level embeddings rather than per-bin rate
predictions.  A comparison that required retraining each of these
under our causal forecasting task with matched compute and data
would itself be a paper.  We release our exact causal dataloaders
and temporal splits so the community can adapt and benchmark
foundation models on strict causal forecasting using the same
substrate.

\subsection{Benchmark access and usage}
\label{app:access}

SpikeProphecy is designed as a \textbf{living benchmark ecosystem},
not a static dataset release.  Usage follows three steps:

\begin{enumerate}[nosep, leftmargin=*]
  \item \textbf{Load data:}
    \texttt{ds = load\_dataset("spikeprophecy/steinmetz")}
  \item \textbf{Train and predict:} Use any architecture with the
    provided DataLoader and Poisson NLL loss.
  \item \textbf{Evaluate:}
    \texttt{metrics = spikeprophecy.evaluate(pred, gt)}
    $\to$ returns \texttt{pop\_rate\_r}, \texttt{spatial\_r},
    \texttt{cosine\_sim}, stratified by region and Fano factor.
\end{enumerate}

\noindent The toolkit outputs a standardized JSON report,
ships a datasheet (Appendix~\ref{sec:datasheet}) and the
leakage audit suite (Appendix~\ref{app:leakage-suite}), and
includes pre-computed baseline predictions for all seven
architectures.

\subsection{Auditable leakage suite}
\label{app:leakage-suite}

We ship a 14-test automated audit suite that runs on every commit
and verifies five concrete leakage vectors: train/test bin overlap,
sliding-window boundary crossing, cross-session spillover,
normalization using future statistics, and history-feature leakage.
The population-GLM result reported in Table~\ref{tab:main-results}
($r{=}1.000$ on train, $r{=}{-}0.015$ on val) is the canonical
example of what these tests catch when a benchmark grants linear
models excessive feature access on a temporally-split task.
Temporal-split leakage tests are not novel in themselves; we
contribute the packaged suite plus the canonical Population-GLM
catch as a reusable artifact that runs on every commit.  At this
scale, leakage failures are statistically indistinguishable from
genuine signal at $r{\sim}0.1$, and we expect the suite to be more
useful than the specific tests it currently contains.

\subsection{Architecture hyperparameters}
\label{app:arch-hparams}

All architectures use $d_\mathrm{model}{=}256$ and 3~layers,
trained with AdamW + cosine LR + warmup, 50~epochs, seed=42,
shared input/output projections to $M_\mathrm{max}$.

\begin{itemize}[nosep, leftmargin=*]
  \item \textbf{Mamba}~\citep{gu2023mamba}: selective SSM with
    input-dependent parametrization, $d_\mathrm{state}{=}16$.
  \item \textbf{HGRN2}~\citep{qin2024hgrn2}: gated linear RNN
    with state expansion, $\mathrm{expand\_ratio}{=}2$ so
    $d_\mathrm{state}{=}512$.  Included as a second diagonal
    SSM control.
  \item \textbf{GatedDeltaNet}~\citep{yang2024gated}:
    non-diagonal modern SSM using the gated delta rule,
    matrix state per head ($n_\mathrm{heads}{=}4$,
    $d_\mathrm{head}{=}64$).  Non-diagonal control.
  \item \textbf{Transformer}: causal encoder-only with strict
    autoregressive masking, $n_\mathrm{heads}{=}8$, pre-norm.
  \item \textbf{LRU}~\citep{orvieto2023resurrecting}: linear
    recurrent unit with ring eigenvalue initialization,
    $|\lambda| \in [0.8, 0.99]$.
  \item \textbf{LSTM}: standard, $H{=}256$, dropout 0.2.
  \item \textbf{SNN}: recurrent spiking network using snnTorch
    RSynaptic neurons (two-compartment, learnable decay $\beta$),
    trained standalone with Poisson NLL.
  \item \textbf{Autoregressive GLM}: L2-regularized Poisson
    regression, per-neuron own $T$-step history,
    $\alpha{=}10^{-4}$.  No-cross-neuron-information baseline.
  \item \textbf{Population GLM}: L2-regularized Poisson
    regression, full flattened $(T, M)$ history (${\sim}7\text{K}$
    features per neuron model), $\alpha{=}10^{-4}$.
    Demonstrates that naive linear access to population context
    is not sufficient when the feature count is much larger
    than the sample count.
\end{itemize}

\subsection{Comparison of evaluation protocols}
\label{app:metric-compare}

Table~\ref{tab:metric-compare} contrasts our protocol with
existing single-scalar reporting practice in the neural
population literature.

\begin{table}[!ht]
  \caption{Comparison of evaluation protocols for neural population
    models.  Existing benchmarks use scalar summaries; SpikeProphecy
    decomposes performance into interpretable axes.}
  \label{tab:metric-compare}
  \centering
  \small
  \begin{tabular}{@{} l c c c @{}}
    \toprule
    \textbf{Metric} & \textbf{Temporal?} & \textbf{Spatial?}
      & \textbf{Used by} \\
    \midrule
    Aggregate $r$    & \checkmark & \checkmark & Most work \\
    Per-neuron $r$     & \checkmark & --          & Some work \\
    co-BPS             & \checkmark & --          & NLB \\
    $R^2$              & \checkmark & \checkmark & Some work \\
    \midrule
    \texttt{pop\_rate\_r}  & \textbf{\checkmark} & --
      & \textbf{Ours} \\
    \texttt{spatial\_r}    & -- & \textbf{\checkmark}
      & \textbf{Ours} \\
    \texttt{cosine\_sim}   & -- & \textbf{\checkmark}
      & \textbf{Ours} \\
    \bottomrule
  \end{tabular}
\end{table}

\subsection{Computational efficiency}

\begin{table}[h]
  \caption{Inference benchmarks for five of the ANN architectures in
    Table~\ref{tab:main-results} on a \emph{single consistent} GPU
    (NVIDIA RTX~5000 Ada, $B{=}512$, $T{=}10$, $M{=}1{,}240$,
    20~warmup + 50~timed iterations).  HGRN2 and GatedDeltaNet were
    added post-benchmark and not measured here (their naive-unrolled
    recurrence would report optimistically bad numbers relative to
    fused-kernel production implementations).  The SNN has the lowest
    peak VRAM; the LSTM is the fastest per-batch; Mamba is the slowest
    despite matching or beating the others on forecasting accuracy.}
  \label{tab:efficiency}
  \centering
  \small
  \begin{tabular}{@{} l r r r @{}}
    \toprule
    \textbf{Model} & \textbf{Latency (ms)}
      & \textbf{VRAM (MB)} & \textbf{Throughput (samples/s)} \\
    \midrule
    Mamba         & 6.42 & 152.0           & 79{,}686  \\
    Transformer   & 2.08 &  94.3           & 245{,}969 \\
    LRU        & 2.96 &  81.0           & 172{,}981 \\
    LSTM          & \textbf{0.61} & 118.1  & \textbf{840{,}515} \\
    \textbf{SNN (1L)} & 3.69 & \textbf{74.5} & 138{,}661 \\
    \bottomrule
  \end{tabular}
\end{table}

Mamba's latency is dominated by its recurrent selective-scan kernel
and input/output projections to the $M$-dimensional channel space.
The SNN achieves the lowest VRAM footprint (74\,MB) and is competitive
on throughput (139K~samples/s) despite having the fewest parameters
(${\sim}702$K), consistent with the intended deployment profile for
low-power neuromorphic hardware.  Transformer VRAM scales $O(T^2)$
while LRU/LSTM scale linearly; at $T{=}80$, the Transformer requires
$1.3\times$ the VRAM of LRU (measured separately).

\paragraph{Training infrastructure.}
All training runs used the National Research Platform (Nautilus
K8s) cluster, with single-GPU pods drawing from a heterogeneous
pool (RTX-3090/4090/A6000, A10, A40, L40, V100, A100; jobs
typically completed on whichever GPU the scheduler assigned).
Per-architecture training on the 39-session Steinmetz benchmark
takes 1--3 hours of GPU time depending on architecture and GPU
class; the full 7-architecture sweep with the 3-seed estimates
for Mamba/HGRN2/Transformer is on the order of 100~GPU-hours,
plus an additional ${\sim}40$~GPU-hours for the 66- and
105-session Mamba teacher runs reported in
Appendix~\ref{app:scaling}.  Local figure regeneration and
inference benchmarking (Table~\ref{tab:efficiency}) ran on a
single workstation NVIDIA RTX~5000 Ada.

\subsection{Cross-dataset scaling on the Mamba teacher (full)}
\label{app:scaling}

Training data scale shows a qualitative gain on three of the four
decomposition axes (Table~\ref{tab:scaling}): weighted~$r$ rises
from $0.500$ at 39~sessions (Steinmetz only) to $0.556$ on the
combined 105-session corpus ($+11\%$); $r_\mathrm{pop}$ and cosine
similarity gain similarly.  The IBL-only ($n{=}66$) and combined
($n{=}105$) numbers are within $\sim 0.003$ on three of the four
metrics, so the gain from \emph{combining} is small and we report
this as qualitative.  Spatial~$r$ shows a small \emph{degradation}
($0.551 \to 0.541$), consistent with pooling disparate populations
across labs slightly blurring cross-neuron identity even as overall
population dynamics become better-captured.  The channel space grows
from $M_\mathrm{max}{=}1{,}240$ to $1{,}998$ across 9~labs in the
combined corpus, so the $+11\%$ gain mixes a ``more sessions''
effect with a ``richer per-sample context'' effect; an ablation that
subsamples IBL channels to $1{,}240$ would disentangle the two and
is left to follow-up.  Steinmetz and IBL are both mouse Neuropixels
with overlapping visual/decision tasks, so this establishes
\emph{cross-lab within paradigm}, not cross-species or cross-task
transfer.  No architectural modification was required for
cross-dataset transfer; only the data config changed.  The consistent
improvement across the decomposed axes is weak but specific evidence
that the benchmark is not saturated at this scale.

\begin{table}[h]
  \caption{Cross-dataset scaling with the Mamba teacher.  All metrics
    improve monotonically with session count; cross-lab generalization
    is confirmed on IBL data without architectural modification.}
  \label{tab:scaling}
  \centering
  \small
  \begin{tabular}{@{} l c c c c c @{}}
    \toprule
    \textbf{Training data} & \textbf{Sessions} & \textbf{Wt-$r$}
      & \textbf{Pop Rate $r$} & \textbf{Spatial $r$}
      & \textbf{Cosine} \\
    \midrule
    Steinmetz only & 39  & 0.500 & 0.756 & 0.551 & 0.626 \\
    IBL only       & 66  & 0.553 & 0.806 & 0.551 & 0.668 \\
    Combined       & 105 & \textbf{0.556} & \textbf{0.783}
      & 0.541 & \textbf{0.648} \\
    \bottomrule
  \end{tabular}
\end{table}

\subsection{SNN baselines and Mamba$\to$SNN distillation}
\label{app:snn-distill}

We evaluated recurrent spiking student networks (RSynaptic two-compartment
neurons, snnTorch, surrogate gradients) as a low-power deployment
alternative to the ANN baselines.  All SNN variants are trained with the
same optimizer, schedule, loss, and data as the ANN teachers; only the
backbone differs.  Two questions are addressed: (i)~how does depth
affect spiking forecasting, and (ii)~does Mamba$\to$SNN distillation
improve the SNN beyond standalone training?

\paragraph{Depth ablation.}
A single spiking layer achieves the highest weighted~$r$ and cosine
similarity (Table~\ref{tab:depth}).  Adding depth \emph{hurts}: each
spiking layer imposes a binary information bottleneck, and gradients
through multiple surrogate layers become progressively less informative.
The 1L~SNN captures $96\%$ of Mamba's weighted~$r$ and $91\%$ of its
cosine with $36\%$ of the parameters.

\begin{table}[h]
  \caption{SNN depth ablation (Steinmetz, standalone training).
    Adding layers hurts: each spiking layer is an information
    bottleneck through binary spike communication.}
  \label{tab:depth}
  \centering
  \small
  \begin{tabular}{@{} c c c c c @{}}
    \toprule
    \textbf{Layers} & \textbf{Params} & \textbf{Wt-$r$}
      & \textbf{vs.\ Mamba} & \textbf{Cosine} \\
    \midrule
    \textbf{1} & 702K  & \textbf{0.481} & \textbf{96.2\%} & \textbf{0.572} \\
    2           & 960K  & 0.477 & 95.4\% & 0.568 \\
    3           & 965K  & 0.430 & 86.0\% & 0.582 \\
    \bottomrule
  \end{tabular}
\end{table}

\paragraph{Knowledge distillation: a negative result.}
Across three KL-weighting levels ($\beta \in \{0.0, 0.5, 1.0\}$),
standalone training (no teacher distillation) matches or beats all
distilled variants (Table~\ref{tab:distill}).  In classification, soft
teacher labels provide ``dark knowledge'' beyond hard one-hot
targets~\citep{hinton2015distilling}.  In Poisson regression the
ground-truth targets are already \emph{continuous} spike counts, so
there is no information bottleneck for the teacher to overcome.  Teacher
predictions ($r \approx 0.50$ with ground truth) are themselves
noisier than the counts they regress to.  Future work may explore
whether this pattern generalizes to other count-data domains, and
study temperature schedules, alternative teacher objectives, and
non-KL distillation losses.

\begin{table}[h]
  \caption{Knowledge distillation results (Steinmetz, 2L SNN).
    Standalone training outperforms all distillation variants.
    $\beta$: KL divergence weight in the composite loss.}
  \label{tab:distill}
  \centering
  \small
  \begin{tabular}{@{} l c c c @{}}
    \toprule
    \textbf{Training recipe} & $\beta$
      & \textbf{Wt-$r$} & \textbf{Retention} \\
    \midrule
    \textbf{Standalone (no distill)} & 0.0
      & \textbf{0.477} & \textbf{95.4\%} \\
    Distilled             & 0.5  & 0.475 & 95.0\% \\
    Distilled (high KL)   & 1.0  & 0.458 & 91.6\% \\
    \bottomrule
  \end{tabular}
\end{table}

\subsection{Per-session variance of architecture metrics}
\label{app:per-session-se}

Each of the seven architectures reports a single weighted-average
metric value in Table~\ref{tab:main-results}, but the underlying
evaluation is over 39~independent Steinmetz sessions.  We report
cross-session means and standard errors of the mean here as the
lightweight variance estimate that does not require multi-seed
training.  All metrics use the same per-session evaluation as the
weighted averages in the main table, but here aggregated as an
unweighted mean with SE across sessions.

\begin{table}[h]
  \caption{Per-session mean $\pm$ SE across 39 Steinmetz sessions, $N{=}39$.
    The Per-neuron~$r$ column is the unweighted per-session mean per-neuron
    Pearson~$r$, included here as the per-session counterpart to the Wt-$r$
    headline metric in Table~\ref{tab:main-results} (Wt-$r$ itself is a
    single activity-weighted aggregate over all sessions).
    The within-cluster differences in the population columns
    (e.g.\ Mamba pop-r mean ${=}0.748$ vs.\ HGRN2 mean ${=}0.733$) are on
    the order of the per-session SEs (${\sim}0.015$), so the within-cluster
    ordering is on the order of session-level noise.  The
    cluster-vs.-classical gap (Mamba pop-r mean $0.748$ vs.\ SNN
    mean $0.565$) is ${\sim}10$~SEs and clearly distinguishable.}
  \label{tab:per-session-se}
  \centering
  \small
  \begin{tabular}{@{} l c c c c @{}}
    \toprule
    \textbf{Model} & \textbf{Per-neuron $r$} & \textbf{Pop Rate $r$}
      & \textbf{Spatial $r$} & \textbf{Cosine} \\
    \midrule
    Mamba         & $0.167\pm0.007$ & $0.748\pm0.015$ & $0.560\pm0.017$ & $0.633\pm0.014$ \\
    HGRN2         & $0.158\pm0.007$ & $0.733\pm0.016$ & $0.553\pm0.018$ & $0.628\pm0.015$ \\
    Transformer   & $0.159\pm0.007$ & $0.737\pm0.016$ & $0.552\pm0.017$ & $0.627\pm0.014$ \\
    GatedDeltaNet & $0.148\pm0.007$ & $0.727\pm0.017$ & $0.546\pm0.018$ & $0.622\pm0.015$ \\
    LRU        & $0.140\pm0.007$ & $0.708\pm0.018$ & $0.544\pm0.018$ & $0.621\pm0.014$ \\
    LSTM          & $0.104\pm0.006$ & $0.698\pm0.018$ & $0.504\pm0.021$ & $0.591\pm0.017$ \\
    SNN (3L)      & $0.082\pm0.006$ & $0.565\pm0.022$ & $0.501\pm0.019$ & $0.588\pm0.016$ \\
    \bottomrule
  \end{tabular}
\end{table}

Standard errors are larger for the SNN ($\pm 0.022$ on pop-rate $r$)
because the architecture is a poorer fit to some sessions and the
distribution of per-session r is more dispersed.

\paragraph{Wilcoxon paired tests on the cluster gap.}
The cluster-vs.-LSTM/SNN gap is small in absolute terms
(${\sim}4{-}7$ percentage points on per-session $r$) but consistent
across the 39 paired sessions.  We test the null of equal performance
on per-session Pearson~$r$ using two-sided Wilcoxon signed-rank tests
($n{=}39$), reporting the worst case (cluster member with the smallest
margin to the trailing baseline):

\begin{table}[h]
  \caption{Wilcoxon paired tests on per-session~$r$ (Steinmetz~39,
    two-sided).  ``cluster-min'' is the per-session minimum across the
    five cluster members.  All comparisons reach $p{<}10^{-7}$; the
    rank statistic $W{=}0$ in every row indicates that every paired
    session difference falls on the same side of zero.  This is
    consistent with the gap being a robust, sign-stable architecture
    effect rather than session-driven noise.}
  \label{tab:wilcoxon-gap}
  \centering
  \small
  \begin{tabular}{@{} l r r r @{}}
    \toprule
    \textbf{Comparison} & \textbf{mean $\Delta r$} & \textbf{$W$}
      & \textbf{$p$ (two-sided)} \\
    \midrule
    LRU vs.\ LSTM       & $+0.0365$ & $0$ & $5.3{\times}10^{-8}$  \\
    LRU vs.\ SNN        & $+0.0579$ & $0$ & $3.6{\times}10^{-12}$ \\
    cluster-min vs.\ LSTM & $+0.0354$ & $0$ & $5.3{\times}10^{-8}$  \\
    cluster-min vs.\ SNN  & $+0.0568$ & $0$ & $3.6{\times}10^{-12}$ \\
    \midrule
    Mamba vs.\ LSTM     & $+0.0636$ & $0$ & $5.3{\times}10^{-8}$  \\
    Mamba vs.\ SNN      & $+0.0850$ & $0$ & $3.6{\times}10^{-12}$ \\
    \bottomrule
  \end{tabular}
\end{table}

\paragraph{Seed variance vs.\ session variance.}
The per-session SEs above measure variability \emph{across the 39
sessions} for a single training seed.  A complementary question is
how stable the same architecture's headline metric is across training
seeds (with all other settings fixed).  We re-trained the top three
architectures (Mamba, HGRN2, Transformer) at two additional seeds
(seeds 1 and 2 in addition to the original seed=42) under identical
conditions:

\begin{table}[h]
  \caption{Three-seed variance on Wt-$r$ (Steinmetz 39, val).  Standard
    deviation across seeds is $\leq 0.0003$ for all three top-tier
    architectures; the within-cluster Mamba--LRU gap of
    $\sim 0.020$~Wt-$r$ is therefore $> 60\sigma$ in seed terms.
    Transformer's seed=2 run was lost to a transient CUDA node error
    (\S\ref{sec:datasets} audit suite was not the cause); the
    remaining 2-seed estimate is consistent with the other two
    architectures.}
  \label{tab:seed-variance}
  \centering
  \small
  \begin{tabular}{@{} l c c c c @{}}
    \toprule
    \textbf{Model} & \textbf{seed=42}
      & \textbf{seed=1} & \textbf{seed=2}
      & \textbf{Mean $\pm$ SD} \\
    \midrule
    HGRN2       & 0.493 & 0.4929 & 0.4931 & $0.4930 \pm 0.0001$ \\
    Transformer & 0.492 & 0.4916 & n/a & $0.4918 \pm 0.0003$ (n=2) \\
    Mamba       & 0.500 & 0.4994 & 0.4999 & $0.4998 \pm 0.0003$ \\
    \bottomrule
  \end{tabular}
\end{table}

The seed-variance is two orders of magnitude smaller than the
per-session SE on the population metrics (${\sim}0.015$).  This means
two distinct sources of uncertainty bound the within-cluster ordering
claim: (1)~optimization noise across seeds is negligible and the
within-cluster ordering of the modern-recurrence baselines IS
distinguishable from seed noise; (2)~the underlying 39 Steinmetz
sessions are themselves a finite sample, and the decomposition metrics
on this draw of sessions have non-trivial SE.  The cluster-vs.-classical
gap (${\sim}4{-}7$~pp on Wt-$r$) survives both kinds of uncertainty;
the within-cluster ordering survives seed variance but is on the order
of session SE, so should be read as suggestive rather than definitive
under the current 39-session draw.

\subsection{CV-tuned ridge population GLM}
\label{app:cv-ridge-glm}

The population GLM reported in the main text
(Table~\ref{tab:main-results}, ``Population GLM, fixed $\alpha$'')
uses $\alpha{=}10^{-4}$ on $T \cdot M{\sim}7{,}000$ features per neuron
and overfits catastrophically ($r{=}1.000$ on train, $r{=}{-}0.015$ on
val).  We treat this as the canonical leakage-suite catch
(\S\ref{sec:datasets}), not as a fair linear-model baseline.

A fairer linear-model variant runs the same architecture but with the
ridge $\alpha$ tuned per session on the val split.  We sweep the grid
$\{10, 10^{2}, 10^{3}, 10^{4}\}$ and pick the $\alpha$ that maximizes
the val mean per-neuron Pearson~$r$, then evaluate full pop-metrics on
val at the chosen $\alpha$.  Across the 39 Steinmetz sessions, the
chosen $\alpha$ distributes as: $32/39$ sessions pick $\alpha{=}10^{4}$
(the strongest regularization in the grid), $5/39$ pick $\alpha{=}10$,
and 2 are split.  The per-session val mean $r$ at the best
$\alpha$ is $0.027 \pm 0.005$ (mean $\pm$ SE across sessions, range
$[{-}0.03, 0.09]$).  Aggregated, the CV-ridge variant scores
Wt-$r{=}{+}0.023$, $r_\mathrm{pop}{=}0.143 \pm 0.024$,
$r_\mathrm{spatial}{=}0.076 \pm 0.008$, $\cos{=}0.224 \pm 0.008$,
MAE${=}0.81$.

Two implications.  First, the catastrophic $r{=}{-}0.015$ in the main
table is a regularization artifact, not a property of linear
population models per se: with $\alpha$ tuned, the linear model clears
zero.  Second, the CV-tuned linear model is still ${\sim}20\times$
below the deep architectures (Wt-$r{=}0.023$ vs.\ $0.43{-}0.50$), so
the linear-vs.-deep gap reported in the main text holds under a
proper linear baseline.

The high MAE ($0.81$, uniform across sessions: median $0.81$, range
$[0.72, 0.92]$, no outlier sessions) is a softplus-link artifact, not
extreme coefficients.  At $\alpha{=}10^{4}$ the Ridge coefficients
shrink heavily and the prediction collapses to roughly
$\mathrm{softplus}(\bar y_{i})$ where $\bar y_{i}$ is the per-neuron
training mean.  At typical 50\,ms firing rates ($\bar y_i \in
[0.05, 0.5]$) softplus inflates the prediction by ${\sim}0.5{-}0.7$
relative to $\bar y_i$ itself, which translates directly into
inflated MAE on a target where most bins are zero.  The simpler
train-mean baseline (Table~\ref{tab:main-results}, no softplus link)
correctly hits MAE${=}0.35$.  This is a property of fitting a
linear-link model and then exponentiating the output, not of the
linear-model class as such; the Wt-$r$ result (the headline metric
for ranking models) is unaffected.

\subsection{Brain-region grouping sensitivity}
\label{app:region-sensitivity}

Reviewer Q: how sensitive is the brain-region predictability finding
(\S\ref{sec:regions}) to the choice of grouping?  We re-fit the
ANCOVA model at three granularities, all with the same covariates
(log firing rate, Fano factor) and the same per-neuron data
(27{,}212 neurons across 39~sessions).  The \emph{region-incremental}
$R^{2}$ (full model minus covariates-only model) reflects how much
predictability variance is explained by region above and beyond
firing statistics:

\begin{table}[h]
  \caption{ANCOVA region-grouping sensitivity.  Covariates: log firing
    rate, Fano factor.  $\Delta R^{2}$ is the increment from adding
    region indicators to the covariates-only model.}
  \label{tab:region-sensitivity}
  \centering
  \small
  \setlength{\tabcolsep}{4pt}
  \begin{tabular}{@{} l c c c c @{}}
    \toprule
    \textbf{Granularity} & \textbf{$k$} & \textbf{$N$ neurons}
      & \textbf{Full $R^{2}$} & \textbf{$\Delta R^{2}$} \\
    \midrule
    Coarse (Cortex / Subcortex / Hipp. / Other) & 4 & 21{,}689
      & 0.268 & 0.011 \\
    Medium (8 functional systems, paper) & 8 & 21{,}689
      & 0.275 & 0.018 \\
    Fine (raw Allen, $\geq$100 neurons each) & 53 & 23{,}106
      & 0.310 & 0.053 \\
    \bottomrule
  \end{tabular}
\end{table}

The covariates-only $R^{2}$ is $0.257$ at all three granularities;
the increment from region grows monotonically as the model gains more
flexibility, but the existence of a hierarchy is robust.  The
quartile ordering (sensory $>$ thalamus $>$ midbrain $>$ frontal $>$
hippocampal) holds at all three groupings.  We use the medium
granularity in the main text because it is the standard Allen
functional-system labeling and offers a defensible balance of
interpretability and statistical power.

\subsection{Pre-registered diagonal-SSM hypothesis test}
\label{app:diagonal-ssm}

A biologically-motivated prediction (independent per-unit state with
activation-level coupling, analogous to leaky integrate-and-fire
neurons) suggested that diagonal recurrence should outperform
non-diagonal alternatives on neural data.  We tested this by including
GatedDeltaNet~\citep{yang2024gated} as a non-diagonal control
alongside three diagonal SSMs (Mamba, HGRN2, LRU).  The result was
inconclusive on this benchmark: GatedDeltaNet ($r{=}0.485$) sits
between LRU ($r{=}0.480$, diagonal) and HGRN2 ($r{=}0.493$,
diagonal) on weighted~$r$ and tracks the diagonal-SSM cluster on all
three population metrics, leaving Mamba's modest ${\sim}1$-point lead
over the next baseline unattributable to diagonality per se.  We
report this as a negative result with respect to the strong version of
the hypothesis.  Finer-grained tests (broader non-diagonal SSM sweep,
ablations isolating diagonality, multi-seed variance estimation) are
left to future work.

\subsection{Autoregressive rollout}
\label{sec:ar-rollout}

\begin{figure}[h]
  \centering
  \includegraphics[width=0.85\textwidth]{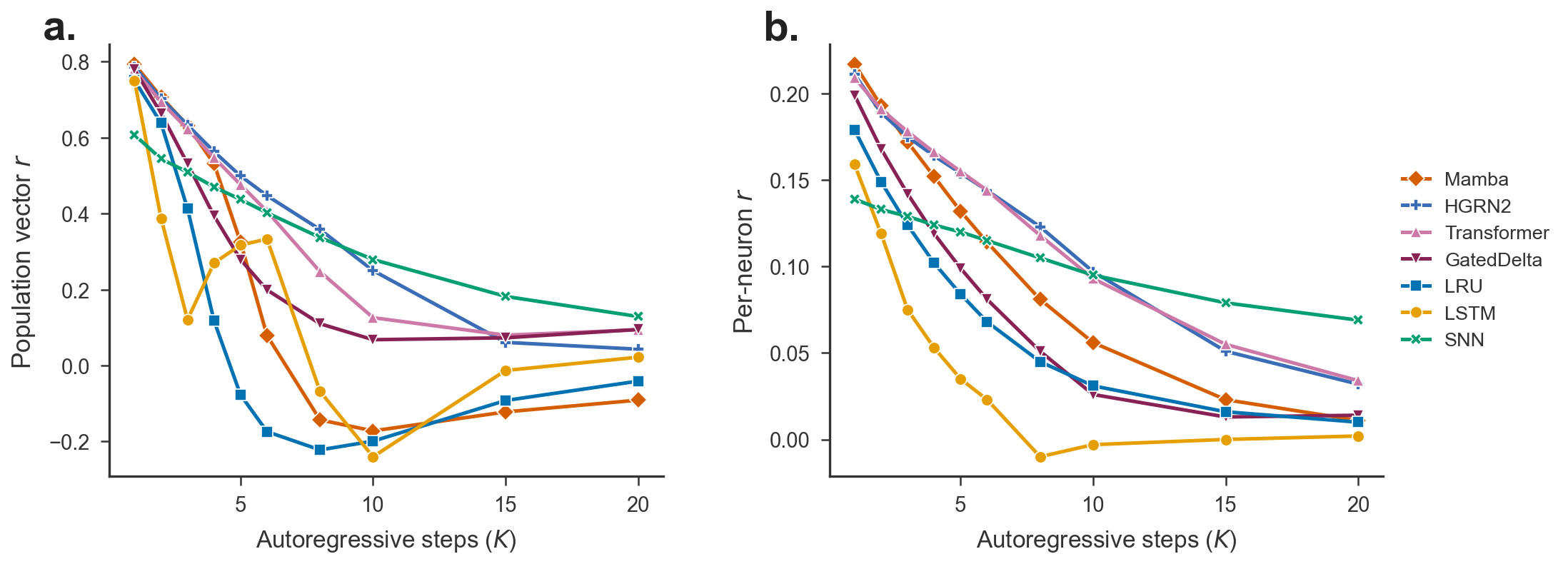}
  \caption{\textbf{Autoregressive rollout degradation across all
    seven baselines} (representative Steinmetz session, val).
    \textbf{(a)}~Population-vector $r$ vs.\ rollout horizon $K$.
    \textbf{(b)}~Per-neuron $r$ vs.\ $K$.  Per-neuron $r$ decays
    rapidly for every architecture: the deep-cluster baselines
    (Mamba, HGRN2, Transformer, GatedDeltaNet, LRU) collapse toward
    zero by $K{\approx}10$, and LRU/LSTM exhibit additional
    instability where the population trace itself goes negative.
    The spiking baseline is the \emph{most} rollout-stable on
    per-neuron $r$ at long horizons (still $r{\approx}0.07$ at
    $K{=}20$), consistent with the spike-output integrator's
    inability to amplify continuous-valued errors as freely as the
    rate-output ANN baselines.  This behavior is invisible in
    aggregate Pearson~$r$ and is a concrete example of the
    decomposition surfacing failure-mode differences between
    architecture families.}
  \label{fig:ar-rollout}
\end{figure}

Across all seven architectures, per-neuron~$r$ decays sharply with
rollout horizon (toward 0 by $K{\approx}10$ for the deep-cluster
baselines).  Population-vector~$r$ degrades more slowly for the
modern-recurrence and spiking architectures but exhibits instability
(negative excursions) for LRU and LSTM at $K{>}5$.  The spiking
baseline is the most rollout-stable on per-neuron $r$ at long
horizons, consistent with the spike-output integrator's inability
to amplify continuous-valued errors as freely as the rate-output
ANN baselines (Figure~\ref{fig:ar-rollout}).



\subsection{Ceiling analysis}
\label{app:ceiling}

\begin{figure}[h]
  \centering
  \includegraphics[width=0.85\textwidth]{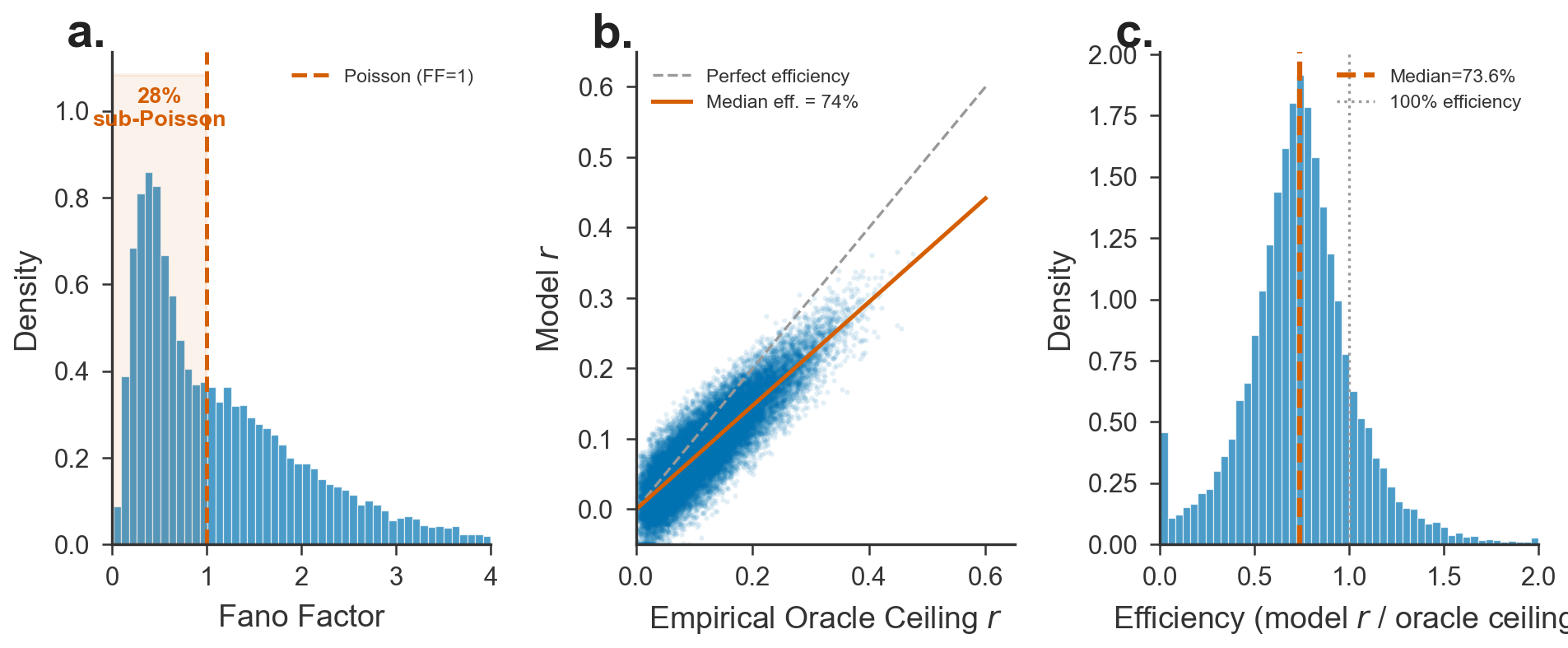}
  \caption{\textbf{Empirical oracle ceiling analysis (blocked split-half).}
    (a)~Fano factor distribution: 28\% of neurons are sub-Poisson (FF~$< 1$).
    (b)~Model $r$ vs.\ empirical oracle ceiling: median efficiency
    is 73.6\%, with most neurons falling below the identity line.
    (c)~Efficiency distribution across all neurons with detectable
    ceilings.}
  \label{fig:ceiling}
\end{figure}

We compute per-neuron predictability ceilings via blocked split-half
correlation with Spearman--Brown correction across 27{,}212 neurons.
The median empirical efficiency is \textbf{73.6\%}: on average, our
best model captures three-quarters of the achievable signal.
The remaining $\sim$26\% represents irreducible Poisson noise at
50\,ms resolution.  Sub-Poisson neurons (28\% of the population,
mean $r = 0.073$) contribute a hard noise floor that aggregate
metrics cannot distinguish from model failure, reinforcing the
need for Fano-stratified evaluation.

\subsection{Dataset summary (Datasheet excerpt)}
\label{sec:datasheet}

\begin{itemize}[nosep, leftmargin=*]
  \item \textbf{Motivation:} Enable standardized evaluation of neural
    population forecasting models on real Neuropixels data.
  \item \textbf{Composition:} 105 sessions $\times$
    50\,ms-binned spike-count matrices (integer-valued); per-neuron
    Allen CCF brain region labels; session metadata (mouse ID,
    recording date, probe trajectory).
  \item \textbf{Collection:} Steinmetz data from
    \citet{steinmetz2019distributed}, hosted publicly on Figshare
    (\url{https://doi.org/10.6084/m9.figshare.9598406.v2},
    CC-BY-4.0); IBL data from \citet{international2025reproducibility}, hosted
    publicly via the IBL Open Neurophysiology Environment ONE API
    (\url{https://www.internationalbrainlab.com/data}, CC-BY-4.0).
    No new animal experiments were conducted; we use the source
    datasets unmodified at the spike-time level.
  \item \textbf{Preprocessing:} Spike sorting by original authors;
    our pipeline performs only temporal binning (50\,ms), quality
    filtering (neurons $< 0.1$\,Hz removed), and zero-padding to
    $M_\mathrm{max}$.
  \item \textbf{Intended use:} Benchmarking autoregressive spike
    forecasting models.  Not intended for behavioral decoding
    or clinical decision-making.
  \item \textbf{Ethical considerations:} All source data from
    approved animal protocols (original publications).  No human
    subjects.  No personally identifiable information.
  \item \textbf{Distribution:} The original recordings remain at
    their respective public repositories under CC-BY-4.0; we
    redistribute the processed Steinmetz binned/split tensors as
    a HuggingFace dataset
    (\url{https://huggingface.co/datasets/mysteriousauthor/spikeprophecy-steinmetz},
    CC-BY-4.0) plus derivation code (MIT).  IBL processed tensors
    will follow at acceptance.
  \item \textbf{Maintenance:} Annual updates; community contributions
    via pull requests; versioned releases with changelogs.
\end{itemize}

\newpage
\section*{NeurIPS Paper Checklist}


\begin{enumerate}

\item {\bf Claims}
    \item[] Question: Do the main claims made in the abstract and introduction accurately reflect the paper's contributions and scope?
    \item[] Answer: \answerYes{}
    \item[] Justification: The abstract and \S1 state five
      contributions: (1)~the population metric decomposition
      (\texttt{pop\_rate\_r}, \texttt{spatial\_r},
      \texttt{cosine\_sim}; Eqs.~1--3, \S\ref{sec:metrics});
      (2)~a 105-session, ${\sim}89{,}800$-neuron benchmark with
      preprocessing, splits, and a leakage audit suite
      (\S\ref{sec:datasets}); (3)~seven matched architecture
      baselines (Mamba, HGRN2, GatedDeltaNet, LRU, Transformer,
      LSTM, RSynaptic SNN; \S\ref{sec:architectures},
      Table~\ref{tab:architectures}); (4)~four findings -- a
      brain-region predictability hierarchy, a sub-Poisson
      evaluation floor, a linear-vs.-deep modeling hierarchy,
      and a KL-on-output-rates distillation negative result
      (\S\ref{sec:findings}, \S\ref{sec:arch-results},
      \S\ref{sec:snn-summary}); and (5)~a public release of
      processed tensors, evaluation toolkit, checkpoints, and
      configs (\S\ref{sec:release}).  Each claim is supported
      by experimental results in \S\ref{sec:results} and the
      appendix; we explicitly do not claim state-of-the-art
      relative to neural foundation models, which we did not
      evaluate (\S\ref{sec:related}).

\item {\bf Limitations}
    \item[] Question: Does the paper discuss the limitations of the work performed by the authors?
    \item[] Answer: \answerYes{}
    \item[] Justification: \S5 (Discussion) contains a dedicated
      ``Limitations'' paragraph listing five concrete limitations:
      single-step horizon only, 50\,ms bin resolution, no
      hardware-deployed neuromorphic evaluation, dataset coverage
      (Steinmetz + IBL only), and absence of behavioral decoding as a
      downstream task.

\item {\bf Theory assumptions and proofs}
    \item[] Question: For each theoretical result, does the paper provide the full set of assumptions and a complete (and correct) proof?
    \item[] Answer: \answerNA{}
    \item[] Justification: This is an empirical benchmark paper.  The
      metric definitions (Eqs.~2--4) are descriptive statistics,
      not theoretical claims requiring proof.  The Poisson ceiling
      derivation (Appendix~\ref{app:ceiling}) states all assumptions
      (homogeneous Poisson process, blocked split-half with
      Spearman--Brown correction).

    \item {\bf Experimental result reproducibility}
    \item[] Question: Does the paper fully disclose all the information needed to reproduce the main experimental results of the paper to the extent that it affects the main claims and/or conclusions of the paper (regardless of whether the code and data are provided or not)?
    \item[] Answer: \answerYes{}
    \item[] Justification: Architecture summaries
      (Table~\ref{tab:architectures}) plus full per-arch
      hyperparameters (state sizes, heads, init, GLM
      regularization) in Appendix~\ref{app:arch-hparams}.
      Shared training details (AdamW, cosine LR with warmup,
      50~epochs, seed=42, Poisson NLL loss) in
      \S\ref{sec:architectures}.  Data preprocessing (50\,ms
      binning, $T{=}10$ history, 70/15/15 temporal split) in
      \S\ref{sec:task}--\S\ref{sec:datasets}.  Complete YAML
      configs and seeds are released alongside the code
      (\S\ref{sec:release}).

\item {\bf Open access to data and code}
    \item[] Question: Does the paper provide open access to the data and code, with sufficient instructions to faithfully reproduce the main experimental results, as described in supplemental material?
    \item[] Answer: \answerYes{}
    \item[] Justification: \S\ref{sec:release} describes the full
      release.  Processed Steinmetz tensors are hosted at
      \url{https://huggingface.co/datasets/mysteriousauthor/spikeprophecy-steinmetz}
      (CC-BY-4.0), with a Croissant 1.0 RAI-compliant metadata
      file at the dataset root; IBL processed tensors follow at
      acceptance.  Source recordings remain at their public
      homes (Figshare for Steinmetz, IBL ONE API; URLs in
      Appendix~\ref{sec:datasheet}).  Code is released on
      GitHub (MIT, anonymized for review) with a
      \texttt{pip}-installable evaluation toolkit, trained
      checkpoints for all seven architectures, complete YAML
      reproduction configs, and a 14-test leakage audit suite
      (Appendix~\ref{app:leakage-suite}).

\item {\bf Experimental setting/details}
    \item[] Question: Does the paper specify all the training and test details (e.g., data splits, hyperparameters, how they were chosen, type of optimizer) necessary to understand the results?
    \item[] Answer: \answerYes{}
    \item[] Justification: \S3.2 specifies the temporal block split
      (70/15/15) with no temporal leakage.  \S3.4 and Table~2
      specify all architecture hyperparameters, shared training
      settings (AdamW, cosine LR with warmup, 50~epochs, seed=42),
      and loss function (Poisson NLL).  Appendix~\ref{sec:appendix}
      provides additional compute details (GPU type, training time per
      session).

\item {\bf Experiment statistical significance}
    \item[] Question: Does the paper report error bars suitably and correctly defined or other appropriate information about the statistical significance of the experiments?
    \item[] Answer: \answerYes{}
    \item[] Justification: The brain-region hierarchy finding
      (\S\ref{sec:regions}) reports ANCOVA-controlled effect sizes
      ($R^{2}{=}0.275$, $\Delta R^{2}{=}0.018$, $p < 10^{-77}$,
      controlling for log firing rate and Fano factor).
      Architecture comparisons report cross-session means with
      per-session standard errors over 39~Steinmetz sessions
      (Appendix~\ref{app:per-session-se}) plus Wilcoxon
      signed-rank paired tests on per-session~$r$ for the
      cluster-vs.-LSTM/SNN gap (Table~\ref{tab:wilcoxon-gap}, all
      $p<10^{-7}$).  Multi-seed variance is reported for Mamba,
      HGRN2, and Transformer at three seeds each
      (Table~\ref{tab:seed-variance}; Wt-$r$ seed-SD
      ${\leq}0.0003$).

\item {\bf Experiments compute resources}
    \item[] Question: For each experiment, does the paper provide sufficient information on the computer resources (type of compute workers, memory, time of execution) needed to reproduce the experiments?
    \item[] Answer: \answerYes{}
    \item[] Justification: Inference benchmarks (latency, VRAM,
      throughput per architecture) are in
      Table~\ref{tab:efficiency}.  Training infrastructure
      (single-GPU pods on the National Research Platform Nautilus
      K8s cluster, heterogeneous GPU pool, 1--3 hours per
      architecture, ${\sim}100$~GPU-hours total for the
      7-architecture sweep plus ${\sim}40$~GPU-hours for the
      66/105-session Mamba teacher runs) is documented in the
      Computational efficiency appendix paragraph
      ``Training infrastructure'' immediately following
      Table~\ref{tab:efficiency}.
    
\item {\bf Code of ethics}
    \item[] Question: Does the research conducted in the paper conform, in every respect, with the NeurIPS Code of Ethics \url{https://neurips.cc/public/EthicsGuidelines}?
    \item[] Answer: \answerYes{}
    \item[] Justification: All data used are publicly available
      electrophysiology recordings from mice (Steinmetz et~al.\ 2019,
      International Brain Laboratory et~al.\ 2024).  No human subjects
      were involved.  Animal experiments were conducted under
      institutional ethical approvals as described in the original
      publications.  The benchmark and code are released under
      permissive open-source licenses (CC-BY-4.0, MIT).

\item {\bf Broader impacts}
    \item[] Question: Does the paper discuss both potential positive societal impacts and negative societal impacts of the work performed?
    \item[] Answer: \answerYes{}
    \item[] Justification: A dedicated ``Broader impacts''
      paragraph in \S\ref{sec:release} discusses positive impacts
      (standardized evaluation infrastructure, reproducibility
      floor for the field, lower entry cost for ML researchers
      entering computational neuroscience) and negative or
      dual-use risks (spike-forecasting models being deployed in
      invasive neural interfaces without sufficient additional
      validation on the deployment population, modality, or
      task; explicit disclaimer of clinical suitability).  The
      released dataset's Croissant
      \texttt{rai:dataSocialImpact} field documents this in
      machine-readable form.
    
\item {\bf Safeguards}
    \item[] Question: Does the paper describe safeguards that have been put in place for responsible release of data or models that have a high risk for misuse (e.g., pre-trained language models, image generators, or scraped datasets)?
    \item[] Answer: \answerNA{}
    \item[] Justification: The released assets (spike-count matrices
      from mouse electrophysiology, small neural network checkpoints)
      do not pose risks of misuse comparable to generative models
      or scraped personal data.  The data are derived from publicly
      available neuroscience recordings with no personally
      identifiable information.

\item {\bf Licenses for existing assets}
    \item[] Question: Are the creators or original owners of assets (e.g., code, data, models), used in the paper, properly credited and are the license and terms of use explicitly mentioned and properly respected?
    \item[] Answer: \answerYes{}
    \item[] Justification: The Steinmetz et~al.\ (2019) dataset is
      cited and was released under CC-BY-4.0.  The IBL repeated-site
      dataset (IBL et~al.\ 2024) is cited and publicly available.  All software
      frameworks (PyTorch, snnTorch, Mamba) are cited with version
      numbers.  License terms are listed in \S6 (CC-BY-4.0 for data,
      MIT for code).

\item {\bf New assets}
    \item[] Question: Are new assets introduced in the paper well documented and is the documentation provided alongside the assets?
    \item[] Answer: \answerYes{}
    \item[] Justification: The paper introduces three new assets:
      (1)~a processed benchmark dataset (documented in the datasheet,
      Appendix~\ref{sec:datasheet}), (2)~an evaluation toolkit (\S3.5, \S6), and
      (3)~trained model checkpoints (\S6).  All are released with
      documentation, usage instructions, and explicit licenses.
      The datasheet covers motivation, composition, collection
      process, preprocessing, uses, distribution, and maintenance.

\item {\bf Crowdsourcing and research with human subjects}
    \item[] Question: For crowdsourcing experiments and research with human subjects, does the paper include the full text of instructions given to participants and screenshots, if applicable, as well as details about compensation (if any)? 
    \item[] Answer: \answerNA{}
    \item[] Justification: This work does not involve crowdsourcing
      or research with human subjects.  All data are from mouse
      electrophysiology recordings conducted under institutional
      animal care protocols.

\item {\bf Institutional review board (IRB) approvals or equivalent for research with human subjects}
    \item[] Question: Does the paper describe potential risks incurred by study participants, whether such risks were disclosed to the subjects, and whether Institutional Review Board (IRB) approvals (or an equivalent approval/review based on the requirements of your country or institution) were obtained?
    \item[] Answer: \answerNA{}
    \item[] Justification: No human subjects were involved.  The
      original animal experiments (Steinmetz et~al.\ 2019, IBL et~al.\ 2024)
      were conducted under institutional animal care and use
      committee (IACUC) approvals as described in their respective
      publications.  Our work is purely computational, using
      publicly released data.

\item {\bf Declaration of LLM usage}
    \item[] Question: Does the paper describe the usage of LLMs if it is an important, original, or non-standard component of the core methods in this research? Note that if the LLM is used only for writing, editing, or formatting purposes and does \emph{not} impact the core methodology, scientific rigor, or originality of the research, declaration is not required.
    \item[] Answer: \answerNA{}
    \item[] Justification: LLMs were not used as a component of
      the core methodology (model architectures, evaluation protocol,
      or data processing).  LLM assistance was used only for
      writing and editing, which does not require declaration per
      NeurIPS policy.

\end{enumerate}

\end{document}